\documentstyle[epsf,fleqn,aps]{revtex}
\begin{document}
\title{Interfaces with a single growth inhomogeneity and 
anchored boundaries}

\author{M. D.  Grynberg}

\address{Departamento de F\'{\i}sica, Universidad Nacional de La Plata,
(1900) La Plata, Argentina.}
\maketitle
\date{\today}
\maketitle

\begin{abstract}
The dynamics of a one dimensional growth model involving 
attachment and detachment of particles is studied 
in the presence of a localized growth inhomogeneity along with
anchored boundary conditions. At large times, the latter 
enforce an equilibrium stationary regime which allows for an
exact calculation of roughening exponents. The stochastic
evolution is related to a spin Hamiltonian whose spectrum gap
embodies the dynamic scaling exponent of late stages. For vanishing
gaps the interface can exhibit a slow morphological transition  
followed by a change of scaling regimes which are studied numerically.
Instead, a faceting dynamics arises for gapful situations.

\vspace{10 pt}


\vspace{-12 pt}
\end{abstract}

\vskip 0.2cm
\vskip2pc

\section{Introduction}

After two decades of investigations the dynamics of growing 
interfaces continues to be a subject of enormous interest, 
providing a framework to compare experiments, simulations 
and theory, let alone the wide range of 
applications encountered \cite{Krug}. 
Despite the vast diversity of morphologies in which growing
surfaces can evolve, most numerical analysis and theoretical 
studies pointed out the onset of scaling regimes  emerging
at both large time and length scales. This enabled a classification 
of apparently dissimilar processes in terms of universality 
classes characterized by a set of scaling exponents which dominate 
the late evolution stages \cite{Krug,Tim}. 

A common feature associated with these processes
is the possible emergence of rather slow temporal crossovers
in which the early dynamics exhibits quite different roughening 
characteristics from those observed in the asymptotic limit 
\cite{Evans}. The presence of growth rate inhomogeneities
or growth defects localized within small spatial regions
of the substrate plane (columnar defects),
is one of the simplest mechanisms whereby such crossovers 
can be observed \cite{Krug,Wolf}. 
Another possibility is realized by anchoring 
conditions through which nonequilibrium fluctuations  
are completely suppressed at the interface boundaries \cite{Krug2}.
The main interest in those situations is in the morphological 
phase transitions that may occur at large times.
In this work we investigate the change of scaling regimes
accompanying these transitions by means of a prototype 
restricted solid on solid (RSOS) growth model \cite{Meakin,Plischke} 
combining both of these mechanisms in one dimension.
As we shall see, anchored boundaries are essential 
for the appearance of equilibrium regimes which in turn allow
for a simple calculation of roughening exponents at late stages.

At the phenomenological level of the Kardar-Parisi-Zhang equation
\cite{KPZ}, thoroughly studied in the continuum theory of kinetic 
roughening, large scale morphology changes  
can be accounted for by assuming a growth velocity 
which is a symmetric function of the local inclination
of the interface. The analysis of this equation under a growth 
rate inhomogeneity then leads to a {\it nonequilibrium} steady 
state (SS) which emerges in the form of a macroscopic hill \cite{Wolf}.
Here, we show that a similar scenario may also arise under equilibrium 
and near equilibrium conditions dominated by unusual scaling regimes.

In studying the latter it is useful to consider the mean square 
fluctuations of the average interface 
height $\bar h (t)$ which yields a measure of the interface {\it width}  
\begin{equation}
\label{width0}
W^2 (L,t) = \frac{1}{L} \,\sum_n \langle \,
\left[\, h_n (t)\, - \,\bar h(t) \,\right]^2\,
\rangle\,,
\end{equation}
where the brackets denote an ensemble average over all possible
evolutions of heights $\{h_n\}$ forming the interface at time t,
which grows on a substrate of size $L$. 
On general grounds it can be argued that $W$ scales as \cite{Family}
\begin{equation}
\label{scaling} 
W (L,t) = L^{\zeta} \,f(t/L^z)\,,
\end{equation}
where the scaling function $f (c)$ satisfies
\begin{equation}
\label{affine}
f (c) \sim \cases{ c^{\,\zeta/z} \:\:\:\:\:\:\:
{\rm for} \:\:\: c \ll 1\,,\cr
{\rm const} \:\:\:\:\:\, {\rm for} \:\:\:c \gg 1\,. \cr}
\end{equation}
Hence, it follows that finite systems saturate as 
$W \propto L^{\zeta}$, whereas in the thermodynamic limit
the asymptotic growth is ruled by the exponent $\beta = \zeta/z\,$,
that is $W \propto t^{\beta}\,$.
The exponent $\zeta\,$ describes the roughness dependence 
of the interface width on the typical substrate size. In turn 
the exponent $z\,$, often known as the dynamic exponent, gives the 
fundamental scaling between length and time.

In practice, Eqs. (\ref{scaling}) and (\ref{affine}) yield a standard
procedure which is often followed to extract and corroborate
scaling exponents predicted by other approaches and 
certainly we will make use of this hypothesis as well.
However, due to the presence of the crossovers referred to above,
a complementary procedure would be needed if the former becomes 
exceptionally slow. To this aim, we will
also exploit the known equivalence between the 
RSOS growth models already mentioned and a one dimensional 
gas of hard core particles undergoing an asymmetric exclusion process
\cite{Krug,Meakin,Plischke} (see Fig. 1). 
The idea is to cast the evolution operator of the associated
master equation of this latter process \cite{Kampen} into a 
suitable quantum spin representation \cite{Mattis,Gunter}
lending itself more readily for a finite size scaling analysis.
Since the  dynamic exponent $z$ of (\ref{scaling}) is ultimately
embodied in the gap behavior of the evolution operator
(or spin ``Hamiltonian''), the study of its lower spectrum can 
then provide information of the late evolution stages in a more 
direct manner. Evidently, this methodology 
along with the evaluation of the roughening exponent $\zeta$ 
-- simplified greatly by the anchoring boundary conditions --
circumvents the problem of arbitrarily long transient regimes
though on the other hand is limited severely by the affordable 
system sizes. A posteriori, it will turn out that already
modest lengths can yield clear finite size trends.
This strategy will be explained in Section II 
and its results compared with those of standard techniques 
given in Section III. We end the paper with Section IV which 
contains our conclusions, along with some remarks on extensions 
of this work.

\section{Spin Representation}

Let us consider the dynamics of lattice aggregation models  
with no overhangs, including both adsorption and desorption of monomers
at random locations of a one dimensional interface 
\cite{Meakin,Plischke}, such as that described in Fig. 1.
As usual, on a coarse grained level of description 
the state of a surface at a given time
is represented by a set of single valued 
functions $h_n (t)$ measuring the surface heights 
at positions  $1 \le n \le L+1\,$ of the growth substrate.
As it was mentioned  above, we are 
interested in boundary conditions 
that suppress completely height fluctuations at $n=1$ and $L+1\,$
for all times, i.e. the interface is anchored at the boundaries.
For simplicity, we study the case where $L$ is even and 
$h_1 = h_{L+1}\,$, whereas deposition and evaporation rates 
$\epsilon\,,\epsilon'\,$ are taken uniformly throughout the
system except on site $\frac{L}{2} + 1\,$ where these probability
values are respectively $\epsilon_0\,,\epsilon'_0\,$.
To prevent the divergence of interface fluctuations in the bulk, 
we impose a RSOS constraint
namely, $\vert h_{n+1} - h_n \vert \equiv 1\,$, $\forall\, n,\,t\,$.
Specifically,  growth (evaporation) events 
$h_n \to h_n + 2\,,$ [$h_n \to h_n - 2\,$], 
with $ n=2, ...\,, L\,,$ occur only at local minima (maxima) 
of the evolving interface. These basic processes and their transition 
rates are schematized in Fig. 1. In turn, the typical configurations 
resulting from these rules at large times are displayed by 
the snapshots of Fig. 2.

It is often convenient to consider the interface slope rather 
than the height, so hereafter we will exploit the known mapping 
between RSOS interface dynamics and quantum spin-$\frac{1}{2}$ systems. 
This correspondence can be easily visualized in the scheme of Fig. 1. 
Associating the height difference $s_n \equiv h_{n+1} - h_n\,$ 
to an eigenvalue of the $z-$component, say, of the Pauli operator 
$\vec\sigma_n\,$ for site $n$, all relevant quantities of the 
interface, such as its width and height-height correlations, 
can be casted in terms of $\frac{1}{2}$-\,spinors. 
By construction, it is clear that the interface heights 
(relative to $h_1\,$) are obtained as $h_n = \sum_{j=1}^{n-1} s_j\,$ 
for $n = 2,\,...\,,L+1\,$. Therefore, the anchoring condition 
$h_1 = h_{L+1}\,$ imposes the vanishing of the total magnetization 
throughout the underlying spin kinetics.

As is well known, the probability distribution of such Markov 
processes is controlled by a master equation \cite{Kampen}
\begin{equation}
\label{ME}
\partial_t P(s,t) = \sum_{s'} \left[\,
R (s' \to s)\, P(s',t) \,- \,R (s \to s')\, P(s,t) \,\right]\,,
\end{equation}
whose transition probability rates 
$R ( s \to s') \in \{ \epsilon, \epsilon' , 
\epsilon_0, \epsilon'_0 \}\,$, now denote
the (biased) spin exchanges at which a generic
configuration $\vert s\,\rangle \equiv
\vert s_1,\,... \,,s_L \rangle\,$ evolves to $\vert s'\rangle\,$
through a single exchange of two consecutive spins.
Starting from a given probability distribution
$\vert P(0) \,\rangle = \sum_s P(s,0)\,\vert s \,\rangle\,$, 
Eq.\,(\ref{ME}) can be conveniently
thought of as a Schr\"odinger like representation in which  
the ensemble averaged state vector $\vert P(t)\,\rangle$
(playing the role of wave function), 
can be evaluated at subsequent times from the action of an evolution
operator (or ``Hamiltonian'') on the initial state, namely
$\vert P (t)\,\rangle = e^{-H \,t} \,\vert P(0)\,\rangle\,$ 
\cite{Mattis}. The specific form of $H\,$ can be readily
obtained by introducing spin-$\frac{1}{2}\,$ raising and 
lowering operators $\sigma^+,\sigma^-\,$, along with spin 
occupation fields $\hat{\rm \bf n} = \sigma^+ \sigma^-\,$.
It is then straightforward to show that the stochastic dynamics 
of Eq.\,(\ref{ME}) is accounted by the  operator
\begin{eqnarray}
\nonumber
H = &-& \sum_{n=1}^{L-1} \,\left(\, \epsilon_n\,\,
\sigma^+_n \,\sigma^-_{n+1} \,+ \,\epsilon'_n\,
\sigma^+_{n+1} \,\sigma^-_n \,\right)
\\
\label{evol0}
&+& \sum_{n=1}^{L-1}\,\left[\,\epsilon_n\,\,\hat {\rm \bf n}_{n+1} \,
( 1 - \hat {\rm \bf n}_n )\, + \,\epsilon'_n \,\hat {\rm \bf n}_n \,
( 1 - \hat {\rm \bf n}_{n+1} )\, \right]\,,
\end{eqnarray}
where $\epsilon_n, \epsilon'_n\,$ are shorthands denoting respectively
\begin{equation}
\label{rates}
\epsilon_n,\, \epsilon'_n = \cases{ \epsilon_0,\epsilon'_0
\:\:\: {\rm for} \:\:\: n = \frac{L}{2}\,,\cr\cr
\epsilon,\epsilon'\:\:\:\:\: {\rm otherwise}\,, }
\end{equation}
while the anchoring condition 
$h_1 = h_{L+1}\,$ confines the dynamics within the subspace 
$\sum_n   \hat {\rm \bf n}_n \equiv~L/2\,$.
The biased hopping terms of (\ref{evol0}) 
clearly describe the original growth--desorption events 
(see Fig. 1), while conservation of probability 
requires in turn the appearance of the remaining (diagonal) 
particle-vacancy correlators. We address the reader to 
Ref. \cite{Gunter} for a more detailed derivation in related systems.

\subsection{Detailed Balance}

Before continuing with an alternative spin representation
more suitable to study dynamical aspects at large times, 
we pause and consider the SS of Eq.\,(\ref{evol0})
along with its implications on the interface character.

Given two spin configurations $\vert s \,\rangle = 
\vert ...\,, s_n,\, -s_n,\,...\,\rangle\,$, 
$\vert s' \,\rangle = \vert ...\,, -s_n,\, s_n,\,...\,\rangle\,$,
differing at most in the state of two neighboring $n, n+1\,$ locations, 
evidently detailed balance probabilities in (\ref{ME})
will hold provided that 
\begin{eqnarray}
\nonumber
P(s) \, \epsilon'_n &=& P(s')\, \epsilon_n\,,\:\:\:\: {\rm if}\:\:\:
\langle s' \vert \,H\, \vert s \,\rangle = - \epsilon'_n\,,\\
P(s) \, \epsilon_n &=& P(s')\, \epsilon'_n\,,\:\:\:\: {\rm if}\:\:\:
\langle s' \vert \,H\, \vert s \,\rangle = - \epsilon_n\,.
\end{eqnarray} 
This can be readily satisfied
defining a hard-core particle (up spin) potential 
\begin{equation}
\label{potential}
V (n) = \sum_{j=1}^n\, \ln \left( 
\frac{\epsilon_j}{\epsilon'_j} \right) = 
n\,\ln \left(\frac{\epsilon}{\epsilon'}\right) \,+ \, 
\ln \left(\frac{\epsilon_0\, \epsilon'}{\epsilon'_0\,\epsilon}\right)\,
\Theta \left( n - L/2 \right)\,,
\end{equation}
through which the equilibrium distribution is simply obtained as
\begin{equation}
\label{distribution}
P (s_1,\,...\,,s_L) \propto \exp -\left[\,
\frac{1}{2}\sum_n V (n)\, (1+s_n)\,\right]\,.
\end{equation}
When $\epsilon = \epsilon'$, these probabilities further enable us to
construct the partition function (normalization constant), 
height profiles (spin densities), as well as
the spin correlation functions needed to derive the equilibrium 
interface width. For $\epsilon \ne \epsilon'$ a rather 
involved recursive relation in the particle number can be 
obtained for all these quantities, but its analytic solution is not
reachable by standard means \cite{Wilf}. However, this
case does {\it not} yield a rough interface as statistical
fluctuations become exponentially suppressed in time
(see Section III B). 

If $\epsilon = \epsilon'$, the step function potential 
(\ref{potential}) permits to divide the system into two independent 
regions $[1,\frac{L}{2}],\, [\frac{L}{2}+1, L]\,$ with  
${\frac{L}{2} \choose m}\,$ configurations having
$\frac{L}{2}-m\,$ and  $m$-particles 
respectively  ($0 \le m \le \frac{L}{2}\,$).
Hence, the partition function normalizing the above
SS distribution is given by
\begin{equation}
\label{partition}
Z = \sum_{m=0}^{\frac{L}{2}}\, {\frac{L}{2} \choose m}^2 r^m\,,
\end{equation}
where $r = \epsilon'_0/\epsilon_0\,$. 
Using analogous arguments, we can also obtain
the reduced partition function $Z_i$ which arises from the
occupation of a given site $i$,
and evaluate
the spin density $\langle \sigma^z_i \rangle = 2\,Z_i/Z - 1\,$.
This results in
\begin{equation}
\label{density}
Z_i = \cases{ \,\sum\limits_{m=0}^{\frac{L}{2} -1}\:
{\frac{L}{2} -1 \choose m}\, {\frac{L}{2} \choose m} \,r^m\,,
\:\:\:\: {\rm for} \:\:\: i \le \frac{L}{2}\,, \cr\cr\
\sum\limits_{m=0}^{\frac{L}{2} -1}\: {\frac{L}{2} \choose m+1}\,
{\frac{L}{2}-1 \choose m} \,r^{m+1},\: {\rm otherwise}  \,,\cr}
\end{equation}
which implies an {\it equilibrium shock} profile stemming
entirely from the inhomogeneous potential (\ref{potential})
at finite particle densities.
In particular,  for $L \to \infty\,$ the analysis of 
Eqs. (\ref{partition}) and (\ref{density}) yields the 
following discontinuity
\begin{equation}
\label{tilt}
\langle \sigma^z_n \rangle = \left[\,1 - 2 \,\Theta(n - L/2) \,\right]\,
\langle s \rangle\,, \:\: \langle s \rangle = 
\frac{ \sqrt{\epsilon_0\phantom{'} } - \sqrt{\epsilon_0'} }
{\sqrt{\epsilon_0\phantom{'} } + \sqrt{\epsilon_0'}}\,.
\end{equation}
Thus, in the hight representation, so long as 
$\epsilon_0'/\epsilon_0 < 1\,$ ($>1\,$)
Eq. (\ref{tilt}) entails the sideways 
growth of a hill (valley) whose sides at large times are tilted by an 
amount of $\pm \langle s \rangle\,$ relative to the substrate.

To determine whether this morphology is actually rough, 
we focus attention on the equilibrium height fluctuations 
$\delta_n \equiv \langle h_n^2 \rangle - \langle h_n \rangle^2\,$.
Consequently, first we evaluate the spin-spin correlations involved in
$\langle \,h_n^2\,\rangle  = (n-1) + 2 \!\!\!\! 
\sum\limits_{i<j\le n-1}\!\!\!\! \langle \, 
\sigma^z_i\,\sigma^z_j\,\rangle\,$.
Thus, once more we recur to the combinatorial reasoning 
and calculate the reduced partition functions  $Z_{i,j}\,$ resulting
from the occupation of two specific sites $i,j\,$. After some 
elementary steps we obtain
\begin{equation}
\label{correlation}
Z_{i,j} = \cases{\,\sum\limits_{m=0}^{\frac{L}{2} - 2}\:
{\frac{L}{2} - 2 \choose m}\, {\frac{L}{2} \choose m}\,r^m, \:\:\:\:
\:\:\:\:\:\:\:\: {\rm for} \:\:\: i < j \le \frac{L}{2}\,,\cr\cr
\,\sum\limits_{m=0}^{\frac{L}{2} - 2}\:
{\frac{L}{2} - 1 \choose m+1}\, {\frac{L}{2} - 1 \choose m}\,r^{m+1},
\:\:\: {\rm for} \:\:\: i \le \frac{L}{2} < j \,,\cr\cr
\sum\limits_{m=0}^{\frac{L}{2} - 2}\:
{\frac{L}{2} \choose m+2}\, {\frac{L}{2} - 2 \choose m}\,r^{m+2},
\:\:\: {\rm othrewise},\cr}
\end{equation}
from which the required spin correlations are computed 
as $\langle \sigma^z_i\, \sigma^z_j \rangle = 4\,Z_{i,j}/Z - 
(\:\langle \sigma^z_i \rangle + \langle \sigma^z_j \rangle + 1\,)\,$.
In the large size limit the analysis of Eqs. (\ref{partition}), 
(\ref{density}) and (\ref{correlation}) 
ultimately yields a rough interface (see leftmost snapshot of Fig. 2), 
whose height fluctuations (canceled at the boundaries), result 
distributed as 
$\delta_n \propto (n\!-\!1)\,(\,1-\, \frac{n\!-\!1}{L}\,)\,$.
This simple form contrast with that observed 
in nonequilibrium SS of anchored self organized interfaces
in which fluctuations in the upper part of the hill are
substantially reduced \cite{Krug2}.

The above correlations can also characterize the saturation
width referred to in Section I. Specifically, 
in the spin representation  it can be easily checked that 
$W^2$ may be rewritten as
\begin{equation}
\label{width}
W^2 (L) = \frac{L^2-1}{6\,L}\,+ \,\frac{2}{L^2}\,
\sum_{i < j }\, i\:(L-j)\:\langle \, \sigma^z_i \sigma^z_j\,\rangle\,.
\end{equation}
In Fig. 3 we display the size dependence of $W$ for several values of 
$\epsilon_0'/\epsilon_0$. It turns out that the roughness exponent 
$\zeta$ bears the discontinuous character of Eqs. (\ref{density}) and 
(\ref{correlation}). 
More precisely,
\begin{equation}
\label{roughness}
\zeta = \cases{ 1\,, \:\:\:\:\:\: {\rm if}\:\:
\epsilon_0' \ne \epsilon_0\,,\cr
1/2\,, \:\: {\rm if}\:\: \epsilon_0' = \epsilon_0\,.\cr}
\end{equation}
Often, a value of $\zeta = 1$ 
is special because it signals that the assumption of a well
defined average orientation of the interface 
(parallel to the substrate plane), becomes inconsistent.
Certainly, this is in line with the tilt 
obtained in Eq. (\ref{tilt}). 
For $\epsilon_0' = \epsilon_0\,$ the conventional (diffusive)
roughening is recovered; here the tilt vanishes and the orientational 
fluctuations at large scales estimated, for example, 
as $W (L,t \to \infty)/L$, decrease with $L$.
We shall revisit this point later on in Section III A.

\subsection{Self Adjoint Representation}

As is known \cite{Kampen}, detailed balance guarantees the 
existence of a representation in which the evolution operator
(\ref{evol0}) is self-adjoint.
Although an exact solution of the (real) $H$-spectrum in the 
thermodynamic limit seems unlikely irrespective of 
its representation, at least a self-adjoint description
can facilitate the numerical analysis of a finite size scaling approach.
Specifically, one can readily find a similarity transformation to map 
(\ref{evol0}) into an hermitian matrix, and thereafter obtain 
the lower eigenmodes dominating the asymptotic kinetics
via recursion-type algorithms, e.g. the Lanczos 
technique \cite{Lanczos}, appropriate to study fair system sizes.

To this aim and with the aid of the particle
potential introduced in (\ref{potential}),
we rotate the above operator around the $z$ spin
direction using a pure imaginary site dependent argument $\varphi (n)$
\begin{equation}
\varphi (n) = \frac{i}{2}\,V (n)\,.
\end{equation}
This rotation is produced by the non-unitary similarity 
transformation $U = e^{-i\,S}\,$ with $S = \frac{1}{2}\,
\sum_n \varphi (n)\,\sigma^z_n\,$, which in turn results in the
direct product
\begin{equation}
\label{similarity}
U = \bigotimes_n\, U_n\,,\:\:\:\:\,
U_n =  \left[\matrix{e^ {\,\frac{1}{4} V (n)} & 
 0 \cr 0  &  e^{\,-\frac{1}{4} V (n)} }\right]\,.
\end{equation}
While the diagonal terms of (\ref{evol0}) remain 
unaltered by $U$, it is straightforward to show that
\begin{equation}
U_n\,\sigma^{\pm}_n\,U^{-1}_n = e^{\mp \,i\,\varphi(n)}\,
\sigma^{\pm}_n\,.
\end{equation}
From this latter transformation, 
one can immediately verify that the rotated (self-adjoint) 
operator ${\cal H} = U \, H \,U^{-1}\,$ becomes an open 
$XXZ$ ferromagnet  with a defect coupling under {\it local} 
magnetic fields, namely
\begin{eqnarray}
\nonumber
{\cal H} = -\frac{1}{2}\, \sum_{n=1}^{L-1} &J_n& \!\!\left[\,
\sigma^x_n\,\sigma^x_{n+1} \,+ \,\sigma^y_n\,\sigma^y_{n+1}\,+\, 
\Delta_n\, ( \sigma^z_n\,\sigma^z_{n+1}\,-\,1)\,\right]\,
\\
\label{evol}
 &-&\, h \,(\,\sigma^z_1 \,-\, \sigma^z_L\,)\,-\,
(\,h_0 - h\,)\,(\,\sigma^z_{\frac{L}{2}} \,- \, 
\sigma^z_{\frac{L}{2}+1}\,)\,.
\end{eqnarray}
where
\begin{eqnarray}
\nonumber
J_n &=&\sqrt{\epsilon_n\,\epsilon'_n}\,,\\
\nonumber
\Delta_n &=& (\epsilon_n+ \epsilon'_n)/
\sqrt{ 4\,\epsilon_n\,\epsilon'_n}\,,\\
h_0 &=& (\epsilon_0\, - \,\epsilon'_0)/4\,,\\
\nonumber
h &=& (\epsilon\, - \,\epsilon')/4\,,
\end{eqnarray}
with $\epsilon_n,\, \epsilon'_n$ taken as in Eq.\,(\ref{rates}).
Thus, we are left with a diagonalization problem which, 
to some extent, is now controllable by standard recursive 
techniques (Section III).

For the sake of completeness it is worth pointing out 
that the similarity transformation (\ref{similarity}) also 
enables us to obtain the SS distribution (\ref{distribution}). 
In fact, exploiting that $H$ is 
a stochastic operator, we can express its {\it left} SS 
$\langle \tilde\psi \vert$
as an equally weighted sum of all accessible configurations 
\cite{Kampen}, i.e. $\langle \tilde\psi \vert \equiv \langle 0 \vert
\sum_{n_1 ... n_{\frac{L}{2}}}
\sigma^-_{n_1}\,...\,\sigma^-_{n_{\frac{L}{2}}}$, where
$\langle 0 \vert$ denotes the ferromagnetic down spin state.
Hence, by construction  $\vert \psi_0 \rangle = U^{-1} \vert 
\tilde \psi \rangle\,$ is a (unnormalized) ground state 
of ${\cal H}\,$, and therefore the SS distribution in the
initial $H$-representation is constructed as
\begin{equation}
U^{-2} \vert \,\tilde \psi \,\rangle \propto \!
\sum_{n_1 ... n_{\frac{L}{2}}}
e^{- V (n_1)} \,...\,\,e^{- V (n_{_{L\!/\!2}})}\:
\sigma^+_{n_1}\,...\:\sigma^+_{n_{\frac{L}{2}}}\vert \,0 \,\rangle\,,
\end{equation}
thus recovering the equilibrium distribution (\ref{distribution}).

Returning to the dynamics, the average value of a diagonal quantity 
$\hat {\cal W}$ --\,such as the ``width operator'' involved in
Eq. (\ref{width})\,--\, varies according to $\langle \tilde\psi \vert \,
\hat {\cal W} \,e^{-H\,t}\,\vert P(0) \rangle\,$ \cite{Gunter}.
Since $\hat {\cal W}$ is invariant under $U$, it is a simple matter 
to check  that in the self-adjoint representation 
$\langle \hat{\cal W} \rangle\,$ reads
\begin{equation}
\label{W}
\langle \hat{\cal W} \rangle (L,t)  = 
\langle\, \psi_0\, \vert \,\hat{\cal W}\,
\vert\,\psi_0\,\rangle \,+\,\sum_{\lambda > 0} \, 
e^{-\lambda \,t}\, \langle\, \psi_0\, \vert \,\hat{\cal W}\,
\vert\,\psi_{\lambda}\, \rangle\,\langle\,\psi_{\lambda}\, 
\vert \, U\,\vert\,P(0)\,\rangle\,,
\end{equation}
where $\{ \vert \psi_{\lambda}\rangle \}$ is a complete orthonormal 
set of eigenstates of  ${\cal H}$ (all with $\lambda \ge 0\,$).
As expected, the role of initial conditions becomes irrelevant
near the equilibrium regime. If the spectrum gap vanishes in the 
thermodynamic limit, the width approach to equilibrium
will involve arbitrarily large times for sufficiently large systems. 
In those situations, finite size scaling analyses of the first excited 
levels $\lambda_L$ would then provide the dynamic $z$-exponent 
ruling over the late roughening stages referred to in Section I.

A distinctive feature arises when all components of the total
angular momentum ${\rm\bf S} = \frac{1}{2} \sum_n \vec\sigma_n\,$
are preserved by ${\cal H}$, namely for $\epsilon = \epsilon'$ 
and $\epsilon_0 = \epsilon_0'\,$. Since $\hat{\cal W}$ just involves 
operators of the form $\sigma_i^z\,\sigma_j^z\,$ 
[\,see Eq.\,(\ref{width})\,], then rather restrictive selection 
rules hold for its matrix elements in (\ref{W}).
Specifically, given that  $\{ \vert \psi_{\lambda}\rangle \}$
can be classified according to the total spin $S$
[\,${\rm\bf S}^2 \equiv S (S+1)\,$], the non-vanishing contributions
to Eq.\,(\ref{W}) come only from states  
$\vert \psi_{\lambda}\rangle$ having
$S = \frac{L}{2}-1$ and $\frac{L}{2}-2\,$ \cite{Alexander}. 
What should be emphasized here is that as soon as 
$[{\cal H},\,{\bf S}\,] \ne 0\,$ the {\it effective} density of states, 
partly responsible for the temporal asymptotic behavior of (\ref{W}) 
when $L \to \infty\,$, is drastically modified as these selection rules 
no longer apply.  We will come back to this issue within the 
numerical context of Section III A.

\section{Numerical Results}

To explore the dynamical consequences of these arguments
we have carried out  Monte Carlo simulations 
as well as finite size scaling analyses of the RSOS model 
referred to above, for a variety of situations. First we focus
attention on the subcase $\epsilon = \epsilon'$ 
where the roles of $\epsilon_0\,$
and $\epsilon'_0$ are clearly interchangeable, so we restrict 
the analysis to, say, $\epsilon'_0/\epsilon_0 \le 1\,$.
In this situation the interface
actually roughens and exhibits two different scaling regimes
accompanying a large scale morphological transition.
The discussion of different bulk probability rates 
is addressed at the second part of the Section. 
It will turn out there that fluctuations decay very rapidly and
prevent the roughening of anchored interfaces at large times.
Instead, a faceting dynamics will emerge regardless of the
inhomogeneity growth rates

\subsection{$\:\:\epsilon = \epsilon'\,$}

Starting from an initially flat configuration
we studied the evolution of the interface width and monitored
the height profiles at different growth stages. 
Fig. 4 displays the width behavior obtained for 
$\epsilon_0'/\epsilon_0 = 0.5\,$ and 0.8.
The rather slow crossover (particularly for 0.8, where 
it is only incipient), deterred us of using larger
substrates, though preliminary simulations averaged over
few evolution samples indicated similar trends. 
The early growing stages support a power law growth 
$W \propto t^{\beta}\,$ extended over more than four decades
with an exponent $\beta \simeq 1/4\,$. This typical diffusive behavior
is accompanied initially by a height profile which is almost
parallel to the substrate, except in the neighborhood of
the growth inhomogeneity. As is shown in the inset of Fig. 4, 
on approaching the asymptopia however, the slopes of the hillsides 
steepen until they reach the equilibrium tilt alluded to in 
Eq.\,(\ref{tilt}). This progressive orientation departure signals
a large scale morphological transition which in turn
is also  reflected in the increase of the growth exponent.
In the late dynamic stages, this can be well fitted by a value of 
$\beta \simeq 1/2\,$ for nearly two decades. 

An alternative determination of this rather peculiar value 
\cite {continuum}, can be implemented
by resorting to the phenomenological scaling assumption referred to in 
Eq.\,(\ref{scaling}) along with the findings and arguments of Section II.
Notice that this provides 
two independent numerical procedures to 
estimate the dynamic $z$-exponent needed for the knowledge of $\beta$.
On one hand, the former can be calculated by studying
the finite size behavior of the first excited levels of ${\cal H}$
(in principle, just the lowest will do), so we diagonalized it exactly
via a recursion type Lanczos algorithm \cite{Lanczos} applied
on the zero magnetization subspace. The huge dimensionality
of this sector, growing as $L \choose L/2\,$, as well as the
lack of translational symmetry of the evolution operator, 
limited our computations to chain lengths of up to 24 spins. 
Nevertheless, they proved to be sufficient for a fair estimation 
of the spectrum gap. We direct the reader to Fig. 5  which suggests 
a decrease $\propto L^{-z}\,$ for the gap and other excited levels 
consistent with a common value of $z = 2\,$. 
On the other hand, 
an independent evaluation of this exponent can be attained by studying
the scaling behavior of the interface width [Eq.\,(\ref{scaling})\,]. 
Fig. 6 exhibits the results of our simulations
for growth substrates of 2500 and 5000 heights
on approaching their saturation regimes. Here, the data collapse
was obtained upon setting $(z,\,\zeta\,) \simeq (2,\,1)\,$, which 
confirms not only the Lanczos estimation but also corroborates 
the roughness exponent quoted in Eq.\,(\ref{roughness}).
Thus, from Eq.\,(\ref{scaling}) it follows that 
the fast roughening behavior already observed  
in Fig. 4, is now recovered by the ratio $\zeta/z\,$.

Since the Lanczos analysis continues to yield values of $z=2\,$ 
holding up to the homogeneous situation (as it should), 
a natural question one can pose is therefore: through which 
feature does the dynamics render a completely different roughening 
behavior at large times as soon as $\epsilon_0 \ne  \epsilon_0'$  ?
In an attempt to provide a plausible explanation
for the appearance of this abrupt change
[see also Eq.\,(\ref{roughness})\,], we resort to the  
observations given by the end of Section II B. The inset of Fig. 5
displays the lower part of the ${\cal H}$-spectrum for 
both inhomogeneous and homogeneous situations. 
As stated above, the latter involves at 
most $L (L-1)/2\,$ contributing levels with $S =L/2 -1,\, L/2 -2\,$, 
while in the former case the sum of Eq.\,(\ref{W}) becomes
much denser as $[{\cal H},\,{\bf S}\,] \ne 0\,$ and an exponential 
number of {\it new} states arises. Of course, the new matrix elements
might eventually change from zero in a continuous manner, but
the density of states (a measure of which is given by the inverse 
of the levels spacing), varies abruptly. Moreover, some
of the low lying excitations controlling the asymptotic regime
($ t \to \infty\,$ holding $t/L^z \ll 1$\,), suggest a rather
narrowly peaked structure which is entirely absent for 
$\epsilon_0 = \epsilon_0'\,$. 

Also, it is interesting to examine whether the large 
scale morphology transition embodied in the slow temporal crossover 
of $W\,$ affects the usual scaling hypothesis of Eqs.\,(\ref{scaling})
and (\ref{affine}).
Thus, we turn to the early dynamic scaling of $W\,$ 
shown by the inset of Fig. 6. In contrast to the super-diffusive 
growth observed at late stages, here the data collapse arises by 
setting standard diffusive exponents $(z,\,\zeta\,) \simeq (2,\,1/2)\,$ 
which in turn yield a scaling function $\propto (t/L^2)^{1/4}\,$, 
in agreement with the early $\beta\,$ exponent measured in Fig. 4.
Hence, combining the late and early scaling regimes it follows that
\begin{equation}
\label{scaling2}
W (L,t,\tau) = L^{\zeta} \,f_{\tau} (t/L^2)\,,
\end{equation}
where $\tau\,$ is a crossover time which depends solely on 
$\epsilon'_0/\epsilon_0\,$ (eventually diverging in the limit 
$\epsilon'_0 \to \epsilon_0\,$), and  $f_{\tau} (c)$
is a universal function defined over three different
scales as
\begin{equation}
f_{\tau} (c) \sim \cases{ c^{\,1/4} \:\:\:\:\:\:\:
{\rm for} \:\:\: c \ll \tau/L^2\,,\:\:\:\:\:\:\:\:\:\:\:\:
(\zeta = 1/2)\,,\cr
c^{\,1/2} \:\:\:\:\:\:\: {\rm for} \:\:\: \tau/L^2 \ll c \ll 1\,,
\:\:\: (\zeta = 1)\,,\cr
{\rm const} \:\:\:\:\: {\rm for} \:\:\:c \gg 1\,. \cr}
\end{equation}

In what follows we address finally to the $\epsilon \ne \epsilon'\,$ 
situation.

\subsection{$\:\:\epsilon \ne \epsilon'\,$}

A quick glance at the evolution of the interface width displayed in
Fig. 7 might render the (wrong) impression that the general case 
$\epsilon \ne \epsilon'\,$, $\epsilon_0 \ne \epsilon_0'\,$ 
bears similar characteristics. 
However this time the growth exponent exhibits an asymptotic
value of $\beta \simeq 3/2,$ which is not understandable
in terms of conventional kinetic roughening theories \cite{continuum}.
In fact, as we shall see below, the interface does not roughen.

Let us first provide a simple explanation for this  
large value of $\beta$ by means of the following heuristic 
considerations. For clarity of argument, assume vanishing
desorption rates and $\epsilon = \epsilon_0$.
Starting from a flat configuration, say with $h_{2k} = 1,\, 
h_{2k-1} = 0\,$, a {\it deterministic} dynamics
arises as $(L-2)/2\,$ deposition attempts  
occur on the initial $(L-2)/2\,$ interface minima. 
Next, we are left with  $h_{2k} = 1,\, h_{2k-1} = 2,\,
h_1 = h_{L+1} \equiv 0\,$ and $(L-4)/2\,$ contiguous minima over which
new $(L-4)/2\,$ depositions will be once again deterministic.
By iterating this argumentation $t$-times,  
this  dynamics leads to a configuration resembling 
a truncated pyramid (see central snapshot of Fig. 2, though for
for a non-deterministic situation, i.e. $\epsilon' \ne 0 \,$).
Specifically, there will be $(L- 2\,t)/2\,$ contiguous minima 
in between $h_{t+2} = t+1\,$ and $h_{L-t-1} = h_{t+2}\,$.
Using the common definition of Eq.\,(\ref{width0}), 
it is then easy to verify that the ``width'' of such 
configuration is simply
\begin{equation} 
\label{heuristic}
W (L,t) = \sqrt{\frac{2\,t^3}{3 \,L}} \,\left[\, 1\, +\, 
{\cal O} \left(t/L\right)\,\right]\,,\:\:\:{\rm i.e.}\: \beta = 3/2\,.
\end{equation}

The above argument describes rather a {\it faceting}
process (terminating at $t \sim L/2\,$), which strictly applies for 
$\epsilon' = 0\,$. Certainly, as soon as $\epsilon' > 0\,$ the dynamics
is no longer deterministic, no matter how small  
$\epsilon'\,$ is. However, for $\epsilon' \ll 1\,$ 
the early desorption attempts become gradually 
unsuccessful as the active region of the interface, i.e.
the number of available minima, decreases inasmuch the sideways
region is increasingly jammed (Fig. 2). Thus, at large times 
a process similar to a faceting dynamics
might be expected, at least for small bulk desorption rates.
In fact, for $\epsilon > \epsilon'\,$ and 
$\epsilon_0 \ge \epsilon_0'\,$ our numerical simulations confirm 
these considerations. No differences were observed between 
sampling histories at large times, so fluctuations 
become asymptotically negligible. 
For $\epsilon_0 < \epsilon_0'\,$ the situation is similar though
another faceting process shows up around the inhomogeneity, 
as displayed by the height profiles in the inset of Fig. 7
(see also rightmost snapshot of Fig. 2).
Of  course, for $\epsilon < \epsilon'$ the roles of 
$\epsilon_0$ and $\epsilon_0'$ are interchanged.
Ultimately, the whole process approaches a non-fluctuating
pile of  slope $\pm 1$, so long as $\epsilon' \ne \epsilon\,$.

To provide an alternative understanding of this fast 
fluctuation decay for generic rate values,
we recur once more to the analysis of the spectrum 
of the evolution operator (\ref{evol}). The gap and levels 
obtained for the sizes within our reach are shown in Fig 8,
but in contrast to the $\epsilon = \epsilon'\,$ situation, here the
finite size trend of these quantities needs further analysis.
To this end, we studied the ${\cal H}$-spectrum within the subspace 
$S^z = L/2 -1\,$ (single spin excitation),
which corresponds to the much simpler anchoring 
case $h_{L+1} - h_1 = L-2\,$. 
For this sector, it is straightforward to check that Eq. (\ref{evol}) 
reduces to the tridiagonal matrix
\begin{equation}
\label{tri} 
{\cal H} = \left[\begin{array}{cccccccc}
\epsilon' & \gamma & 0 & \cdots &  &    & \cdots & 0 \\
\gamma  & \epsilon + \epsilon'  & \gamma & \ddots & & &  & \vdots \\
0 & \ddots & \ddots & \ddots &  &  & &  \\
\vdots & \ddots & \gamma & \epsilon + \epsilon'_0 
& \gamma_{_0} & 0 & &    \\
&   & 0  & \gamma_{_0} & \epsilon' + \epsilon_0 & \gamma & 
\phantom{\ddots} &  \vdots   \\
&  &  &  & \ddots & \ddots & \ddots &  0 \\
\vdots  & &  &  &\ddots &  \gamma  & \epsilon + \epsilon' & \gamma \\
0 & \cdots &  \phantom{\ddots}& & \cdots  & 0  & \gamma & \epsilon
\end{array}\right]\,,
\end{equation}
where 
\begin{equation}
\gamma = - \sqrt {\epsilon \,\epsilon'} \,, \:\:
\gamma_{_0} = - \sqrt {\epsilon_0 \,\epsilon'_0}\,.
\end{equation}
Evidently, this just constitutes the simplest approximation 
to the many body problem of $S^z = 0\,$.
Nevertheless, the comparisons of Fig. 8 indicate that the 
eigenvalues of (\ref{tri}) yet provide an excellent estimation of 
the actual gap and other excited levels obtained through the 
Lanczos scheme. (It is worth pointing out in passing that 
an excellent fit of these quantities was also found 
for $\epsilon = \epsilon'\,$). Using the spectrum gap of the 
single spin approximation, the inset of Fig. 8 strongly 
suggests that the {\it same} gap will persist for $\lambda_1$ in the 
thermodynamic limit of (\ref{evol}). 
Similar gapful results were obtained for other values
of $\epsilon_0'/\epsilon_0\,$. Thus,  fluctuations 
would be suppressed at large times, which is in line with 
the almost invariant values of $W$ observed over many sample 
histories. 

Also, at the gap level the density of states diverges as $L$
in the simplified version of the problem.
For $S^z = 0\,$ however, one might conjecture a much stronger
divergence, probably growing like $\sim e^L\,$, as the number of 
levels between two single excitations tends 
to increase exponentially with the system size (at least
for the small lengths at hand). This would leave us with 
a saturation time $\propto L\,$ in Eq.\,(\ref{W}), which on the
other hand would be in agreement with the termination time
of the faceting process idealized above.
In fact, the numerical simulations displayed in Fig. 9 
lend further support to these speculations.
Clearly, these results exhibit both a saturation time $\propto L\,$, 
as well as an asymptotic scaling regime consistent with 
Eq.\,(\ref{heuristic}), i.e. $W/L \propto (t/L)^{3/2}\,$,
during which both height and width fluctuations are absent. 

In contrast, at early stages the interface displays typical
roughening features. Specifically, the inset of Fig. 9 
exhibits a diffusive scaling regime $W/\sqrt L = f (t/L^2)\,$, 
in turn corroborated by the growth exponent $\beta \simeq 1/4\,$
obtained in larger systems (Fig. 7). This strong departure 
from the faceting description occurs on temporal scales smaller
than a crossover time which turns out to decrease
when $\epsilon'/\epsilon \to 0\,$
(irrespective of $\epsilon_0'/\epsilon_0\,$), 
but eventually diverging in the limit $\epsilon' \to \epsilon \,$, 
$\epsilon'_0 \to \epsilon_0 \,$.

\section{Conclusions}

We have analyzed the characteristics of both early and asymptotic 
dynamics of one dimensional anchored interfaces under 
a growth inhomogeneity. 
There are two  sets of results related respectively 
to equal or different growth-evaporation rates in the bulk. 

For $\epsilon = \epsilon'$, even the slightest departure
from the homogeneous $\epsilon_0 = \epsilon_0'$ situation
is able to produce finite interface tilts as well as huge temporal 
crossovers. The problems posed by the latter have been bypassed
studying {\it separately} the roughness $\zeta\,$ and dynamic 
$z\,$ exponents, the evaluation of which was significantly 
simplified by the appearance of equilibrium SS.
In analyzing finite size scaling trends of the spectrum gap
via the Lanczos method, we found no changes with respect to the 
homogeneous situation, i.e. $z \equiv 2\,$. Consequently, it was argued
that the breaking of the full rotational invariance of the
evolution operator is ultimately responsible for the emergence
of a much heavier density of states accounting both for 
the discontinuity of the roughness exponent $\zeta\,$
(or alternatively, for a different asymptotic growth exponent 
$\beta\,$), as well as for the rise of a new scaling regime 
at large times. 
These expectations were confronted independently
with standard numerical simulations monitoring the evolution
of the interface profile and width. 
At early stages, the latter exhibits a diffusive scaling
regime having basically a non-tilted profile
(except in the inhomogeneity neighborhood), but progressively
approaching a final regime consistent with our scaling exponents
$\zeta = 1,\, z = 2\,$.
However for $\epsilon_0 \to \epsilon_0'\,$, in practice this new
regime might occur at a time so large as to render it 
numerically unobservable.

For $\epsilon \ne \epsilon'\,$ the situation is entirely 
different. Here, the spectrum gap does not vanish in the
thermodynamic limit regardless of the inhomogeneity rates,
and fluctuations between evolution histories at large times 
become negligible. This confirms an heuristic description 
(in turn, tested independently by simulations), 
suggesting that the asymptotic dynamics becomes almost 
deterministic. We may also think of a
synchronous discrete time process in which a randomly chosen
finite fraction, or possibly all of the growth sites, 
are simultaneously updated in a single time step. One 
characteristic feature of such synchronous models is the
occurrence of faceting transitions at large times \cite{Krug},
which also turned out to be the case here.
In contrast, at early stages the interface actually roughens
following a typical diffusive pattern accompanied by a standard
scaling regime.

The analysis of nonequilibrium asymptotic situations, even 
for $d=1\,$, might become rather involved.
In this sense, it will be interesting to elucidate
whether a direct evaluation of $\zeta\,$ could be achieved
using the matrix approach to the asymmetric exclusion process
\cite{Derrida} with both injection and ejection of particles at the 
boundaries, including  one or more hoping defects
(that is, unanchored boundaries and growth inhomogeneities 
in the height representation). 

Higher dimensional extensions of this study would be clearly 
desirable and more realistic. However, the analysis 
of the corresponding quantum spin analogy should 
involve a projector operator to discard all those spin 
configurations having magnetic loops, i.e. 
$\sum_{\bf r} \sigma^z_{\bf r} \ne 0\,$, with ${\bf r}$ on 
a given closed path. Otherwise, the mapping would no longer
represent an interface. The issue as to whether or not
such ideas are actually practical in $d > 1\,$ remains quite open.

\vskip 0.2cm
The author acknowledges support of CONICET, Argentina.

\newpage

\twocolumn

\newpage
$\phantom{}$
\vskip -2cm
\begin{figure}
\hbox{%
\hspace{-0.8cm}
\epsfxsize=4.2in

\epsffile{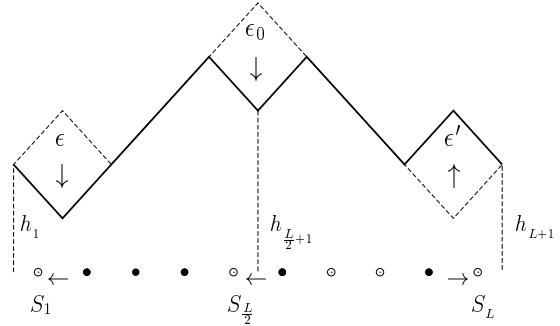}}

\vspace{-6cm}

\caption{Schematic representation of monomer
deposition-evaporation onto a 
RSOS interface with anchored boundaries at $h_1 = h_{L+1}$
The equivalent  spin-$\frac{1}{2}$ ($s_n \equiv h_{n+1} - h_n\,$), 
or hard core particle dynamics involves a left (right) particle hopping  
with rate $\epsilon$ ($\epsilon'$) for monomer adsorption (desorption). 
The corresponding rates for the inhomogeneity
at $h_{\frac{L}{2} +1}$ are $\epsilon_0$ and $\epsilon_0'$.}
\end{figure}

$\phantom{}$
\vskip -3cm
\begin{figure}
\hbox{%
\hspace{-1cm}
\epsfxsize=4in

\epsffile{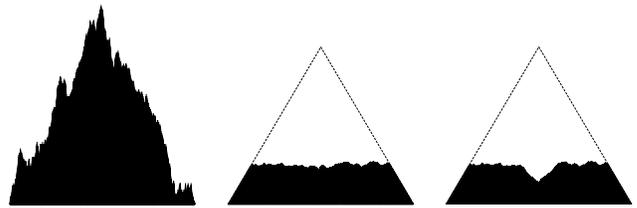}}

\vspace{-6cm}

\caption{Possible evolution scenarios.
Typical snapshots for $L=10^3$ using
$\epsilon'/\epsilon = 1$ with $\epsilon_0'/\epsilon_0 = 0.5\,$
after $t = 10^6$ steps per height (left); 
$\epsilon'/\epsilon = \epsilon_0'/\epsilon_0 = 0.5$ 
at $t=500$ (center); and $\epsilon'/\epsilon = 0.5$ with  
$\epsilon_0'/\epsilon_0 = 5$ at $t=500\,$ (right).
For $\epsilon \ne \epsilon'$, fluctuations are 
progressively reduced on their way to the pile 
configuration denoted by dotted lines with slopes  $\pm 1\,$.}

\end{figure}

\newpage
$\phantom{}$
\vskip -5cm
\begin{figure}
\hbox{%
\epsfxsize=4.3in
\hspace{-1.5cm}

\epsffile{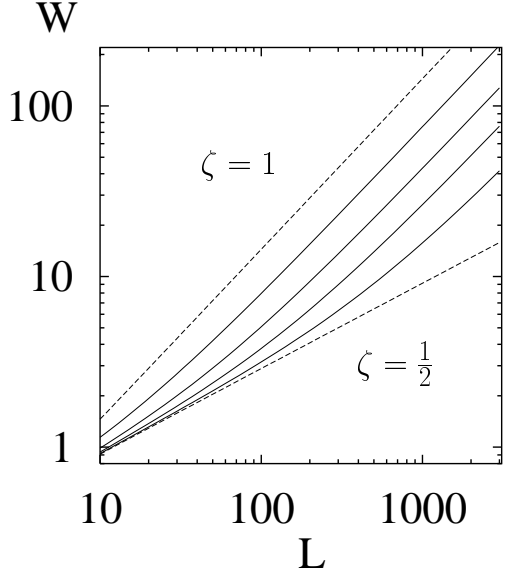}}

\vspace{-3.6cm}

\caption{Equilibrium interface width for $\epsilon/\epsilon' = 1\,$.
Upper and lower dashed curves denote respectively the cases
$\epsilon'_0/\epsilon_0 = \,$ 0 and 1, whereas 
solid lines going downwards stand for 
$\epsilon'_0/\epsilon_0 =  0.1,\,0.3,\, 0.5\,$ and $0.7\,$.}
\end{figure}

$\phantom{}$
\vskip -2.5cm
\begin{figure}
\hbox{%
\epsfxsize=4.3in
\hspace{-1.5cm}

\epsffile{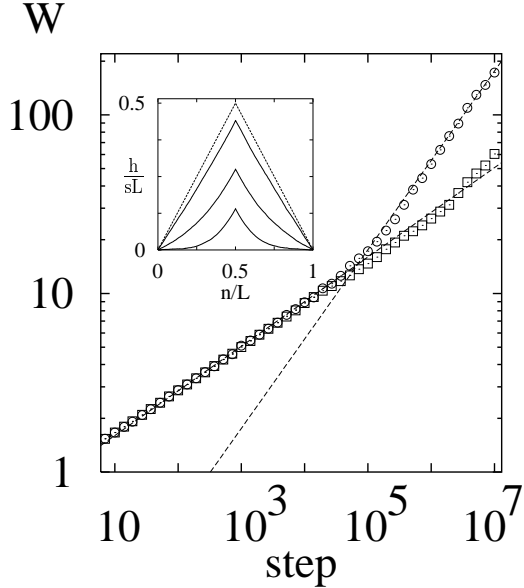}}

\vspace{-5cm}

\caption{Growth of interface width for 
$\epsilon/\epsilon'= 1\,$ using $\epsilon'_0/\epsilon_0 = 0.5\,$ 
(circles) and $\epsilon'_0/\epsilon_0 = 0.8\,$ (squares
showing an incipient asymptotic deviation) 
for $L= 10^4\,$ averaged over 200 histories. The early and 
late slopes of dashed lines, are respectively 
$\beta \simeq \,$ 1/4 and 1/2. 
The inset displays the height profile evolution
for $\epsilon'_0/\epsilon_0 = 0.5\,$
averaged over 2000 histories with $L=1000$,
at $t = 1.5 \times 10^4,\, 6 \times 10^4,\, 2 \times 10^5\,$ 
and $10^6$ [dotted line at the top following closely the tilt 
$s$ of Eq. (\ref{tilt})\,].}
\end{figure}

\newpage
$\phantom{}$
\vskip -5.5cm
\begin{figure}
\hbox{%
\epsfxsize=4.4in
\hspace{-2.3cm}

\epsffile{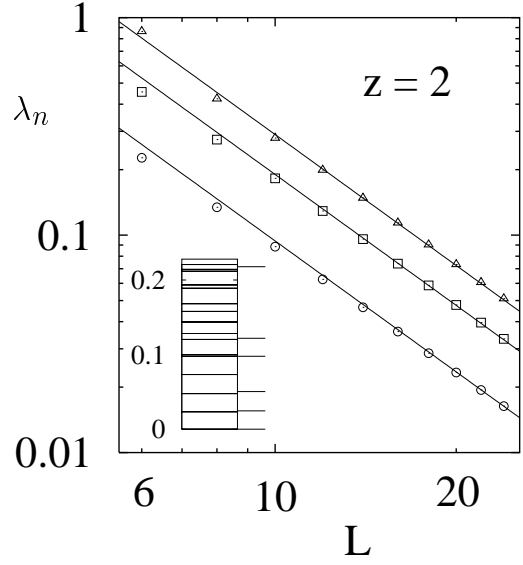}}

\vspace{-3cm}

\caption{Finite size behavior of three lowest excited
levels of the evolution operator (\ref{evol0})
for $\epsilon/\epsilon' = 1\,$ and $\epsilon'_0/\epsilon_0 = 0.5\,$. 
Solid lines have slopes $z = 2\,$.  The inset
compares the lower spectrum of this case ($L = 20\,$, 
10 spin-excitations, framed at the left), with the effective 
levels of the regular situation 
($\epsilon_0/\epsilon'_0 = 1,\, L = 20\,$, 2 spin-excitations).}
\end{figure}

$\phantom{}$
\vskip -4.5cm
\begin{figure}
\hbox{%
\epsfxsize=4.5in
\hspace{-2cm}

\epsffile{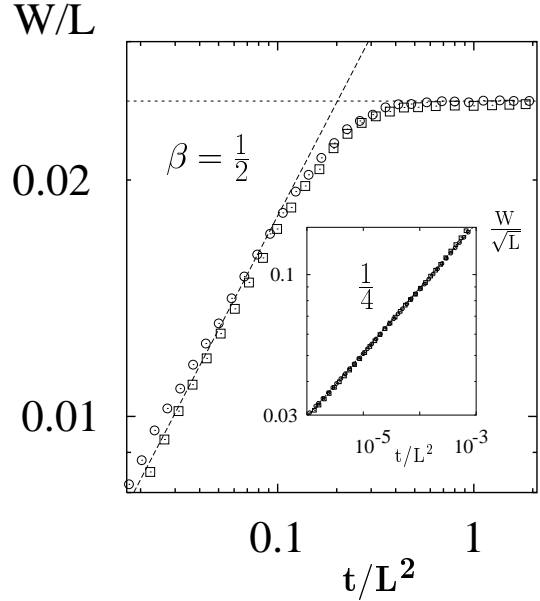}}

\vspace{-4cm}

\caption{Asymptotic finite size scaling regime of the interface
width averaged over 200 histories
using $\epsilon/\epsilon' = 1\,$, 
$\epsilon'_0/\epsilon_0 = 0.5\,$, for $L = 5000\,$ (squares), 
and $L = 2500\,$ (circles).
The saturation values (horizontal line), coincide with those of
Eqs.\,(\ref{correlation}) and (\ref{width}).
The dashed line is fitted with the slope calculated from
the data of Figs. 3 and 5 (i.e. $\beta = \zeta/z\,$).
The inset exhibits an early scaling regime which behaves
diffusively ($\beta = 1/4\,$), alike the initial data of Fig. 4} 
\end{figure}

\newpage
$\phantom{}$
\vskip -4.7cm
\begin{figure}
\hbox{%
\epsfxsize=4.3in
\hspace{-1cm}

\epsffile{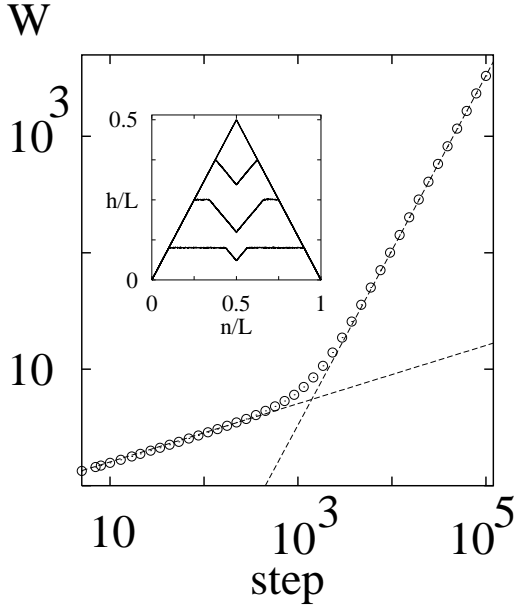}}

\vspace{-4.2cm}

\caption{Evolution of the interface width for 
$\epsilon'/\epsilon =0.5\,$, $\epsilon_0'/\epsilon_0 = 5\,$
and $L = 10^6\,$.  The initial
and final slopes of dashed lines are $\beta = 1/4\,$ and
3/2 respectively. From top to bottom the inset exhibits the
profile of $10^4\,$ heights at $t = 5 \times 10^4,\, 2 \times 10^4,\,
10^4\,$ and $4 \times 10^3\,$. Both width and height fluctuations 
become negligible at large times. At the width level, results of 
different $\epsilon_0'/\epsilon_0$ values closely follow each other
in all evolution stages. In turn, for $\epsilon_0' \le \epsilon_0$
the early profile has no tilt around the inhomogeneity.}
\end{figure}

$\phantom{}$
\vskip -4.5cm
\begin{figure}
\hbox{%
\epsfxsize=4.4in
\hspace{-2.3cm}

\epsffile{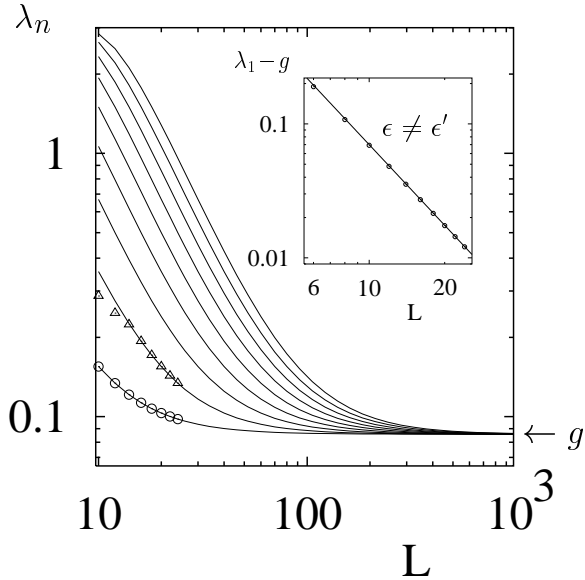}}

\vspace{-4cm}

\caption{Finite size behavior of lower excited levels
in the single spin approximation ($S^z = L/2 -1\,$) of operator
(\ref{tri}) for $\epsilon'/\epsilon = \epsilon'_0/\epsilon_0 = 1/2\,$
(solid lines). Circles and triangles stand  respectively for levels 
$\lambda_1\,$ and $\lambda_2\,$ of Hamiltonian
(\ref{evol0}) with $S^z = 0\,$. 
Above $\lambda_2\,$, further collective excitations (not shown)
appear between successive solid lines. The inset suggests a power 
law convergence of $\lambda_1 (L)\,$ (circles, $S^z = 0\,$), 
towards the gap $g$ obtained in the main panel.}

\end{figure}

\newpage
$\phantom{}$
\vskip -5cm
\begin{figure}
\hbox{%
\epsfxsize=4.5in
\hspace{-2cm}

\epsffile{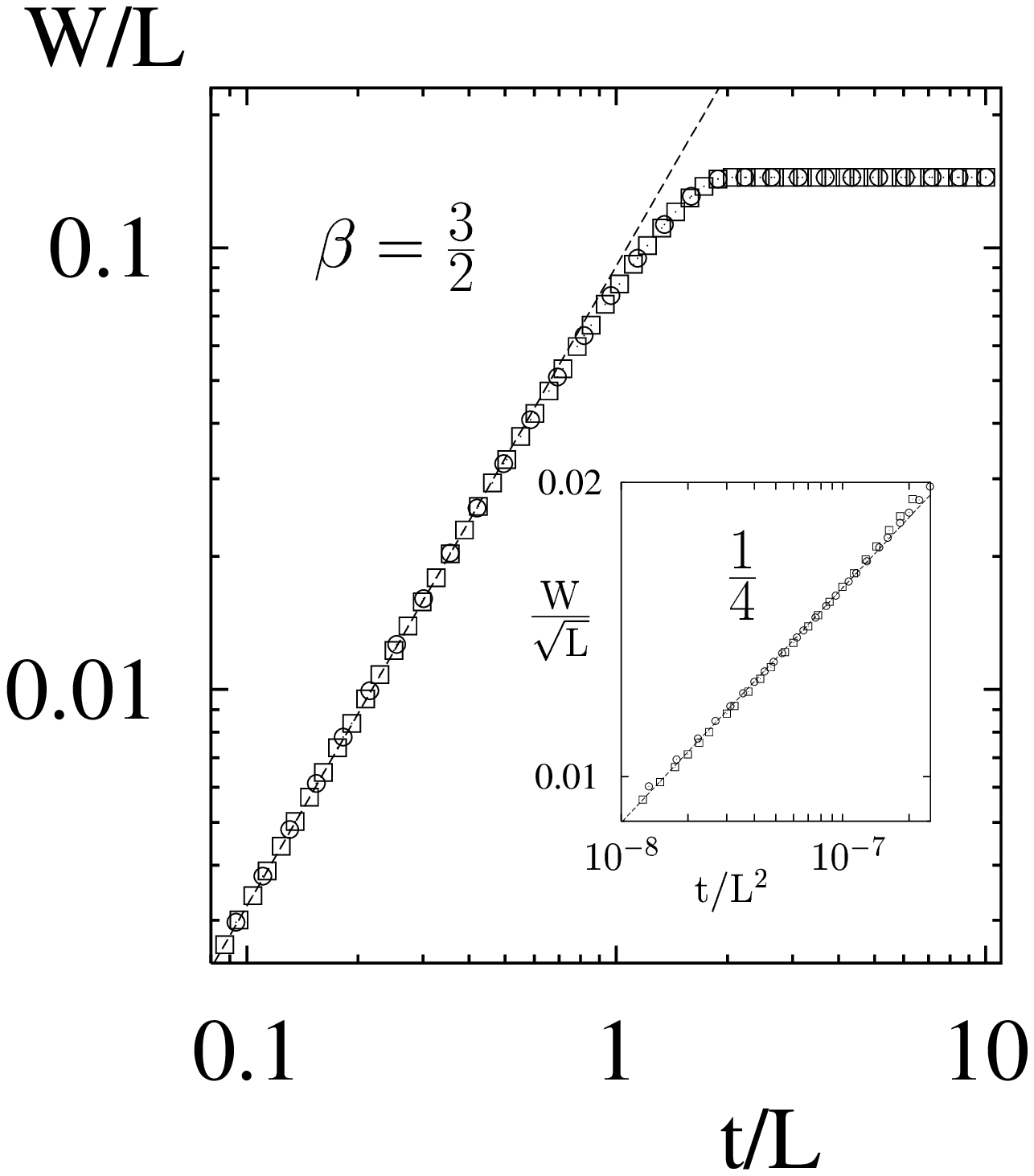}}

\vspace{-4.2cm}

\caption{Late scaling regime of $W$ for $\epsilon'/\epsilon =
\epsilon_0/\epsilon_0' = 0.5\,$ using $L=10^4$ (circles) and
$L=2\times10^4\,$ (squares). The dashed line slope and
scaling form are both consistent with the
faceting process conjectured in (\ref{heuristic}), whereas
differences between sample histories become gradually negligible.
In contrast, the inset results, averaged over 200 histories,
indicate an early scaling regime which is typically diffusive 
$(\zeta = 1/2,\, z = 2,\: \beta = \zeta/z $).}
\end{figure}


\begin{thebibliography}{99}

\bibitem{Krug} For comprehensive reviews and literature list
consult J. Krug, Adv. Phys. {\bf 46}, 139 (1997); 
T. Halpin-Healy and Y.-C. Zhang, Phys. Rep. {\bf 254}, 215 (1995); 
P. Meakin, Phys. Rep. {\bf 235}, 189 (1993);
J. Krug and H. Spohn in {\it Solids far from Equilibrium:
Growth, Morphology and Defects}, edited by C. Godr\`eche, 
(Cambridge University Press, 1992).

\bibitem{Tim} However, the universality issue has been steadily 
brought into question, particularly in higher dimensions. For 
instance, see T. J. Newman and M. R. Swift, Phys. Rev. Lett. 
{\bf 79}, 2261 (1997).

\bibitem{Evans} M. C. Bartelt and J. W. Evans, J. Phys. A {\bf 26},
2743 (1993); B. D. Lubachevsky, V. Privman and S. C. Roy, Phys. Rev. 
E {\bf 47}, 48 (1993); H. C. Kang and J. W. Evans, Surf. Sci. 
{\bf 271}, 321 (1992).

\bibitem{Wolf} D. E. Wolf and L.-H. Tang, Phys. Rev. Lett.
{\bf 65}, 1591 (1990).

\bibitem{Krug2} J. Krug, J. E. S Socolar and G. Grinsten, 
Phys. Rev. A {\bf 46}, R4479 (1992); J. Krug and J. E. S. Socolar
Phys. Rev. Lett. {\bf 68}, 722 (1992); J. Krug and L.-H. Tang
Phys. Rev. E {\bf} 50, 104 (1994).

\bibitem{Meakin} P. Meakin, P. Ramanlal, L. M. Sander and 
R. C. Ball, Phys. Rev. A {\bf 34}, 5091 (1986).

\bibitem{Plischke} M. Plischke, Z. R\'acz and D. Liu, 
Phys. Rev. B {\bf 35}, 3485 (1987).

\bibitem{KPZ} M. Kardar, G. Parisi and Y.-C. Zhang, 
Phys. Rev. Lett {\bf 56}, 889 (1986).

\bibitem{Family} F. Family and T. Vicsek, J. Phys. A {\bf 18},
L75 (1985). However, for an account on anomalous scaling see
J. M. L\'opez, Phys. Rev. Lett. {\bf 83}, 4594 (1999).

\bibitem{Kampen} N.G. van Kampen, {\it Stochastic Processes in 
Physics and Chemistry}, 2nd ed. (North Holland, Amsterdam, 1992).

\bibitem{Mattis} This strategy in related systems
can be traced back over more  than two decades. 
For a review see  D. C. Mattis and 
M. L. Glasser, Rev. Mod. Phys. {\bf 70}, 979 (1998).

\bibitem{Gunter} G. M. Sch\"utz, in {\it Phase Transitions and
Critical Phenomena}, C. Domb and J. L. Lebowitz eds. (Academic, 
London 2000); M. D. Grynberg and R. B. Stinchcombe, Phys. Rev. 
{\bf E} 61, 324 (2000).

\bibitem{Wilf} H. S. Wilf, {\it Generatingfunctionology},
Academic, (1994).

\bibitem{Lanczos} For an account of this time honored technique
see for example, G. H. Golub and C. F. van Loan, {\it Matrix
Computations}, 3rd. ed. (Johns Hopkins University Press, 
Baltimore, 1996).

\bibitem{Alexander} S. Alexander and T. Holstein, 
Phys. Rev. B {\bf 18}, 301 (1978); A. R. Edmonds
{\it Angular Momentum in Quantum Mechanics}, 2nd ed. (Princeton
University Press, 1960).

\bibitem{continuum} In the continuum description of growth equations, 
$\beta = 1/2$ has been conjectured as the upper random 
deposition limit of stochastic roughening. See Ref. [1].

\bibitem{Derrida} For a review, consult B. Derrida and M. Evans
in {\it Nonequilibrium Statistical Mechanics in One Dimension}, 
edited by V. Privman, (Cambridge University Press, 1996); 
see also R. B. Stinchcombe and G. M. Sch\"utz, 
Phys. Rev. Lett. {\bf 75}, 140 (1995).

\end{thebibliography}
\end{document}